# Surface structure and solidification morphology of aluminum nanoclusters

F.L. Tang[*,1,2], W.J. Lu[1], G.B. Chen[1], Y. Xie[2], and W.Y. Yu[2]

1. *State Key Laboratory of Gansu Advanced Non-ferrous Metal Materials, Lanzhou University of Technology，Lanzhou 730050，China*
2. *Key Laboratory of Non-ferrous Metal Alloys and Processing of Ministry of Education，Lanzhou University of Technology，Lanzhou 730050，China*

**ABSTRACT**

Classical molecular dynamics simulation with embedded atom method potential had been performed to investigate the surface structure and solidification morphology of aluminum nanoclusters $Al_n$ ($n$ = 256, 604, 1220 and 2048). It is found that Al cluster surfaces are comprised of (111) and (001) crystal planes. (110) crystal plane is not found on Al cluster surfaces in our simulation. On the surfaces of smaller Al clusters ($n$ = 256 and 604), (111) crystal planes are dominant. On larger Al clusters ($n$ = 1220 and 2048), (111) planes are still dominant but (001) planes can not be neglected. Atomic density on cluster (111)/(001) surface is smaller/larger than the corresponding value on bulk surface. Computational analysis on total surface area and surface energies indicates that the total surface energy of an ideal Al nanocluster has the minimum value when (001) planes occupy 25% of the total surface area. We predict that a melted Al cluster will be a truncated octahedron after equilibrium solidification.

[*]Corresponding author. Email address: tfl03@mails.tsinghua.edu.cn (F.L. Tang).

## 1. Introduction

Aluminum nanoclusters have attracted much attention for their rich display of interesting basic-physics problems and possible applications [1-4]. For many research fields and technological applications (catalysis, cluster deposition [5], microelectronics [6], and superconductivity [7,8]), atomic distribution on their surfaces play a fundamental role, where the surface structure and its quality are of primary importance.

Al nanoclusters had been investigated experimentally and theoretically. Electronic structure evolution of aluminum nanocluster $Al_n$ ($n$ = 1-162) was detected with photoelectron spectroscopy. A possible geometrical packing effect in large clusters ($n > 100$) was found [4]. Ultrasensitive thin-film differential scanning method demonstrated that the melting point of the Al clusters is lower than the value for bulk Al. And the melting point of the clusters is size dependent, decreasing by as much as 140 °C for 2 nm clusters [9]. By means of *in situ* high-temperature scanning tunnelling microscopy, the thermal stability of highly ordered artificial Al nanocluster arrays on vicinal Si (111)-7×7 surfaces had been investigated. It was found that Al nanocluster crystals are stable below 500 °C. Above 500 °C, the Al nanocluster crystal transforms into a surface phase with singular triangle shapes [10].

*Ab initio* density functional pseudopotential technique provided the surface stress and surface energy of Al nanoclutters (0.9-2.0 J/m$^2$) [11] which can be compared with calculated bulk surface energies [12-14]. Using Monte-Carlo simulation with Gupta potentials, Werner explored melting and evaporation transitions in small Al clusters. Dissociation in melting transition can induce a second and possibly much larger local maximum in the specific heat at higher temperature [15]. Icosahedral Al clusters were studied by means of tight binding zero temperature calculations and molecular dynamics simulations. It was found that the free Al clusters have lower melting temperatures and bulk moduli than the bulk material, while they exhibit enhanced low and high-energy phonon density of states [16].

Although literatures had given plenty results focusing on the internal structure and properties to understand aluminum nanoclusters [1-16], there appears to be little published data for atomic distribution on their surfaces and solidification morphologies (especially for larger Al nanoclusters). In this paper, we used classical molecular dynamics simulation with embedded atom method potential to study the details of surface structure and solidification morphology of Al$_n$ ($n$ = 256, 604, 1220 and 2048) nanoclusters.

## 2. Simulation method

The crystal structure of a bulk material at a given temperature and pressure can be predicted by minimizing its free energy [17]. Our approach is to adjust the cell volume and atomic positions until the net pressure or stress is zero. Calculating the free energy at a given volume and then recalculating it after making a small adjustment to the cell volume determines the pressure. During the iterative procedure, a constant volume energy minimization is performed. Hence, each time the cell volume is modified; all atomic positions are adjusted so that they remain at a potential energy minimum. Thus the crystal structure at a given temperature and pressure can be predicted. In this work, atomistic simulation technique in the frame of embedded atom method (EAM) is used to calculate the free energy of aluminum bulk, surfaces and clusters.

In EAM, the cohesive energy of an assembly of $N$ atoms is defined as [18,19]

$$E_{coh} = \sum_i F_i(\rho_i) + \sum_{i>j} \phi(r_{ij}) \text{ and} \qquad (1)$$

$$\rho_i = \sum_{j \neq i} f(r_{ij}), \qquad (2)$$

where $E_{coh}$ is the total cohesive energy, $\rho_i$ is the host electron density at the location of atom $i$ introduced by all other atoms, $f(r_{ij})$ is the electronic density of atom $i$ as a function from its center, $r_{ij}$ is the separation between $i$ and $j$ atoms, $F_i(\rho_i)$ is the embedding energy to embed atom $i$ in the electron density $\rho_i$. $\phi(r_{ij})$ is the Buckingham pairwise potential energy function between $i$ and $j$ Al atoms:

$$\phi(r) = A\exp(-r/\rho) - Cr^{-6}, \qquad (3)$$

where $A$, $\rho$, and $C$ are fitting parameters.

This technique has been used for simulation of many kinds of materials [17,20-27]. Details of this technique are available in [17] and [26]. The potential parameters used for aluminum [18,19] can well reproduce the experimental crystal structure. The calculated aluminum lattice constant $a$ is 4.0479 Å (4.05 Å [28]). The bulk module is 81.34 GPa (76.2 GPa [29]). In addition, simulated melting point of Al is 1000 K (933 K [30]). To simulate melting point, perfect lattice is used whereas there are different types of lattice defects (surface, grain boundary, vacancy etc.) in real bulk Al before melting. This is the reason why the simulated melting point is slightly larger than experimental value. These values by molecular dynamics [31,32] and EAM [33] potential are in a good agreement with the relevant experimental values (in the brackets) and give us confidence to simulate Al surfaces and clusters.

## 3. Results and discussion

### *3.1. Surface energies of aluminum bulk surfaces*

We simulated (001), (111) and (110) surfaces of aluminum using lattice statistics method. To obtain a suitable surface slab model and make the calculations most efficient, the unit cell of Al was extended to two times along the *a*, *b* axis directions and six times along the *c* axis direction. The surface slab has two-dimensional periodic boundary conditions parallel to the surface. The slab was split into two regions (I and II). Above region I, there is a semi-infinite vacuum. During the simulation, the atoms of the region I structural units were relaxed explicitly until there is zero force on each of them, whilst those in region II were kept fixed to reproduce the potential of the bulk lattice on region I. Details of this technique for surface simulation are available in [34]. The lattice parameters *a* and *b* of the slab were kept fixed during the simulation, thus the surface energy $E_s$ can be calculated [27,34] as

$$E_s = \frac{E_{slab} - mE_{bulk}}{S}, \tag{4}$$

where $E_{slab}$ is the total energy of the two-dimensional slab with *m* Al formula units, $E_{bulk}$ is the total energy per unit of the Al bulk and *S* is the surface area of the slab.

In our simulation, six atomic layers in the surface (region I) were relaxed. Atomic displacements of the surface layers are shown in Fig. 1. It is found that the surface atoms have small relaxations compared with their bulk positions. For these three types of surfaces, all the surface layers shift inwards with displacements perpendicular to the surface. And no displacement parallel to the surface takes place. For every kind of surface, the top-layer has the largest displacement: $D_{111} = 1.0\%$, $D_{001} = 1.6\%$ and $D_{110} =$ 2.3% of Al lattice constant $a$ (4.05 Å [28]) for (111), (001) and (110) surface, respectively. In (111) surface, the displacements are almost zero apart from that on the first layer. In (001) and (110) surfaces, the displacements decrease rapidly from the first to the third layer, then decrease slowly and saturate to almost zero at the sixth layer. Hence, it is enough that six surface layers are relaxed to compute surface energies.

Our simulated surface energies of (001), (111) and (110) Al surfaces are shown in table 1 and compared with other computational values. It is found that unrelaxed surfaces have slightly larger surface energies than those of relaxed surfaces, indicating that these surfaces undergo small surface relaxations (as shown above). Relaxed (111) surface has the smallest surface energy $E_{111} = 0.75$ J/m$^2$, whereas relaxed (110) has the largest surface energy $E_{110} = 0.91$ J/m$^2$. Relaxed (001) surface has the middle value ($E_{001} = 0.84$ J/m$^2$). This indicates that (111) surface is more energetically favorable than (001). (110) surface is energetically most unfavorable. It could be found that the order of surface energies ($E_{111} < E_{001} < E_{110}$) of these three surfaces is same as the order of the displacements ($D_{111} < D_{001} < D_{110}$) on their top-layer, just as mentioned in the previous paragraph (Fig. 1). Hence, we suggest that Al surface energy is mainly affected by the first atomic layer of the surface. In addition, atomic densities of the surfaces (the number of atoms per unit area of the surface) can be calculated from the atomic distributions illustrated in Fig. 1: 0.141 atom/Å$^2$, 0.122 atom/Å$^2$ and 0.086 atom/Å$^2$ for (111), (001) and (110) crystal plane, respectively. Obviously, if a surface has a larger atomic density, it has a smaller surface energy. Our simulated surface energies are quantitatively in good agreement with other simulated values (table 1) [12-14]. Most important, our surface energy order is same as that from other simulated results. This gives us confidence to discuss Al nanocluster surface structure with our calculated surface energies.

*3.2. Surface structure of Al nanoclusters*

In order to simulate the Al clusters by molecular dynamics, we obtained four cubic $Al_n$ clusters ($n$ = 256, 604, 1220 and 2048) from the coordinates of three-dimension periodic Al superlattice. We then heated them from 0 K to 1000 K in 40 ps with the time step 1 fs, and kept the temperature for 20 ps. The cubic clusters changed into spherical shape at 1000 K. The melted clusters were cooled down from 1000 K to 300 K in cooling time $t_c$ = 20 to 150 ps with the same time step. It is found that the resulting structures of $Al_n$ clusters depend on the cooling rate. Taking the Al cluster with $n$ = 1220 as an example, we show its Al-Al radial distribution function (RDF) with different cooling time in Fig. 2 and compare the RDF with that of Al lattice/liquid. When $t_c$ < 50 ps, this Al cluster has similar RDF with that of Al liquid. When $t_c$ = 120 and 150 ps, Al cluster has similar RDF peaks with those of Al perfect lattice. Because the resulting structures of Al clusters do not change much more when $t_c \geqslant 120$, we cooled $Al_n$ clusters ($n$ = 256, 604, 1220 and 2048) down slowly from 1000 K to 300 K in 150 ps, and the temperature was kept for 50 ps to collect data.

Figure 3a shows the surface structure of the Al cluster containing 256 atoms. Almost its entire surface is covered with (111) crystal plane atoms. It has two sites covered with (001) crystal plane atoms: one has six atoms (at A site) and the other has four atoms (on the back of the cluster in Fig. 3a). We compared Al-Al bond lengths on cluster (111) surface (Fig. 3a, at B site) with those on bulk (111) surface (Fig. 1). It is found that average Al-Al bond length (2.921 Å) on cluster (111) surface (Fig. 3b) is larger than that of the bulk (111) surface (2.862 Å). We also compared Al-Al bond lengths on cluster (001) surface (Fig. 3a, at A site) with those on bulk (001) surface (Fig. 1). It is found that average Al-Al bond length (2.804 Å) on cluster (001) surface (Fig. 3c) is smaller than the value of bulk (001) surface (2.862 Å). This indicates that cluster (111)/(001) surface has a smaller/larger atomic density and suggests a larger/smaller surface energy compared with the corresponding value on bulk surface.

Figure 4a shows the surface structure of a cluster containing 604 atoms. It has three sites covered with (001) crystal plane atoms: one site has ten atoms (at A site) and every one of other sites has about four atoms (on the back side). The (001) plane at A site is surrounded by four (111) planes. Fig. 4b shows the surface structure of a cluster containing 1220 atoms. This Al cluster has six sites covered with (001) crystal plane atoms: every one of them has about fifteen atoms (for example, B site in Fig. 4b). The surface structure of the cluster containing 2048 atoms is illustrated in Fig. 5. It has six sites are covered with (001) crystal plane atoms: two of them are smaller (every one of them has about twenty atoms, for example, A and B sites in Fig. 5) and the other four sites are larger (on the back side, every one of them has thirty atoms).

On the Al cluster surfaces considered, it is found that both the (111) surface planes and the (001) planes are not perfect. The (111) surface at C site in Fig. 5a is shown in Fig. 5b, which distorts itself to fit the cluster's surface curvature. The (001) surface at B site in Fig. 5a is shown in Fig. 5c, which bends itself to fit the cluster's surface curvature. There are some surface vacancies on Al clusters (near B site in Fig. 4b and at D site in Fig. 5a).

From Figs. 3 to 5, we can find that Al nanocluster surfaces are comprised of (111) and (001) crystal planes. No (110) plane is found on Al nanoclusters in our simulation. The surface area of (001) planes increase as the number of atoms in the cluster increases. (111) crystal planes are predominant on smaller Al cluster surfaces ($n = 256$ and 604). On larger Al clusters ($n = 1220$ and 2048), (111) planes are still dominant but (001) planes can not be neglected.

*3.3. Solidification morphology of Al nanoclusters*

Based on our simulated surface energies and surface structures, we depict Al cluster solidification morphologies in Fig. 6. If an ideal cluster is just covered with {111} perfect planes, it will be an octahedron with eight {111}-equilateral triangles (Fig. 5a). If it is covered with both {111} and {001} perfect planes, it will be a truncated octahedron with fourteen surfaces: six {001}-rectangles and eight {111}-hexagons (Fig. 6b). In Fig. 4a, Al$_n$ ($n = 604$) nanocluster shows a similar morphology to a truncated octahedron: a (001) plane surrounded by four {111} planes. In the truncated octahedron, we denote the surface area of the six {001}-rectangles as $S_{001}$, the surface area of eight {111}-hexagons as $S_{111}$ and the total surface area as $S_{tot} = S_{111}+S_{001}$. The occupancy percentage of {001} planes on the surface can be defined as $P_{001} = 100 \times S_{001}/S_{tot}$ (%). In Fig. 6c, we illustrate the relationship among $P_{001}$, $S_{tot}$ and total surface energy $E_{tot} = E_{001} \times S_{001} + E_{111} \times S_{111}$ (the volume of the truncated octahedron is kept 29000 Å$^3$ unchanged; Al$_{2048}$ cluster has such a volume). When $P_{001}$ increases from zero to 40%, $S_{tot}$ on the truncated octahedron decreases to its minimum value. As $P_{001}$ increases from 40% to 50%, $S_{tot}$ increases.

Generally, $E_{tot}$ will decrease as $S_{tot}$ decreases, as indicated by the arrows in the insert of Fig. 6c. However, note that {001} planes have a larger surface energy $E_{001}$ (0.84 J/m$^2$ = 0.0524 eV/Å$^2$) than that of {111} planes $E_{111}$ (0.75 J/m$^2$ = 0.0468 eV/Å$^2$). If total surface energy decrease introduced by $S_{tot}$'s decrease is over compensated by the energy increase introduced by $S_{001}$'s increase, $E_{tot}$ will increase. When $P_{001}$ increases form 0 to 25%, both $S_{tot}$ and $E_{tot}$ decrease. When $P_{001}$ increases form 25% to 40%, $E_{tot}$ increases though $S_{tot}$ decreases, that is, $E_{tot}$'s decrease introduced by $S_{tot}$'s decrease is over compensated by the energy increase introduced by $S_{001}$'s increase. When $P_{001}$ increases form 40% to 50%, both $S_{tot}$ and $E_{tot}$ increase. Hence, our calculation indicates that ideal Al clusters have the minimum $E_{tot}$ when (001) planes occupy 25% of the total surface area (Fig. 6c).

On the larger Al$_n$ nanoclusters ($n$ = 1220 and 2048), the (001) plane occupancy percentage $P_{001}$ is about 20%. But in smaller Al$_n$ clusters ($n$ = 256 and 604), $P_{001}$ is about 5%, which is much smaller than the value 25% predicted by above computational analysis. We suspect that the surface structure on smaller clusters is controlled not only by surface area and surface energies but also by their internal structure and the total cohesive energy. Our simulations predict that the aluminum clusters will be formed in truncated-octahedral shape after equilibrium solidification, especially for larger Al$_n$ clusters ($n$ > 10$^3$). Ref. [3] discussed at length how the geometry of metallic clusters (mostly noble metals) changes with the cluster size from icosahedral to decahedral up to truncated octahedron by increasing the cluster size in the range explored here for Al. It was reported that Al clusters with stacking faults are also obtained in global optimization on a face-centered cubic lattice with all possible (111) stacking faults allowed, in the case of different model potentials.

We noted that the simulated geometries of Aluminum clusters (Figs. 3-5) have deviations from the ideal geometries predicted by our calculation (Fig. 6b) or by the Wulff construction [3] with crystalline surface energies. During our simulation, Newton-Raphson method and BFGS scheme (Broyden, Fletcher, Goldfarb, Shanno [17]) were used to find the minimum energy structure for Al clusters. To be more precise, this will typically be a local minimum on the global potential energy surface. Trying to locate the global energy minimum is a far more time-consuming and challenging task and one that has no guarantee of success, except for the simplest possible cases [18,19]. From Figs 3-5, we can found that the Al clusters' surfaces are distorted or bended, and Al clusters are quasispherical, just as indicated in Ref. [3].

## 4. Conclusion

Using embedded atom method potential, we performed molecular dynamics simulation on $Al_n$ ($n$ = 256, 604, 1220 and 2048) nanoclusters to investigate their surface structure and solidification morphology. Specific conclusions are as follows:

(1) Al cluster surfaces are comprised of (111) and (001) crystal planes. For smaller Al clusters ($n$ = 256 and 604), (111) crystal planes are dominant on their surfaces. For larger Al clusters ($n$ = 1220 and 2048), (111) planes are still dominant but (001) planes can not be neglected. (110) crystal plane is not found on $Al_n$ cluster surfaces in our simulation.

(2) Computational analysis on total surface area and surface energies indicates that the total surface energy of an ideal Al nanocluster has the minimum value when (001) planes occupy 25% of the total surface area and (111) planes occupy the left. We predict that an Al cluster will adopt truncated-octahedral shape after equilibrium solidification, especially for larger Al$_n$ clusters ($n > 10^3$).

(3) Cluster (111)/(001) surface has smaller/larger atomic density compared with that of bulk (111)/(001) surface.

**Acknowledgment**

The authors would like to thank financial support by LUT Research Development Funding (BS01200905) and National Science Foundation Plan Program. This work was mainly performed in Gansu Province Supercomputer Center with Y.L. Shen and J.W. Zhe's technical help.

Table 1

Surface energies (J/m$^2$) of (001), (111) and (110) crystal planes compared with other calculated values.

|  | Miller index | | |
|---|---|---|---|
|  | (001) | (111) | (110) |
| Unrelaxed | 0.85 | 0.76 | 0.93 |
| Relexed | 0.84 | 0.75 | 0.91 |
| [12] | 0.86 | / | 1.10 |
| [13] | 0.92 | 0.89 | 1.02 |
| [14] | 0.98 | 0.93 | / |

# Figure captions

Fig. 1. (Color online) Atomic displacements perpendicular to Al bulk (001), (111) and (110) surfaces (in percent of Al lattice constant $a = 4.05$ Å). Negative values stand for displacements from the surface to the bulk. $N_{Al}$ is the number of atomic layer from the surface to the bulk. Purple balls show Al atom distributions on the surfaces, and their bond lengths are in units of Å.

Fig. 2. Al-Al RDF with different cooling time ($t_c$ = 20, 40, 120, 150 ps) and RDF of Al lattice/liquid.

Fig. 3. (Color online) Surface structure on Al$_n$ ($n$ = 256) cluster (a): Al-Al bond lengths of (111) surface around B site (b) and Al-Al bond lengths of (001) surface around A site (c). The bond lengths are in units of Å.

Fig. 4. (Color online) Surface structure of Al cluster with 604 atoms (a) and 1220 atoms (b).

Fig. 5. (Color online) Surface structure of Al cluster with 2048 atoms (a): distorted (111) surface plane at C site (b) and bended (001) surface plane at B site (c).

Fig. 6. Solidification morphologies for ideal Al clusters: (a) determined by {111} planes, (b) determined by {111} and {001} planes, and the relationship among total surface area $S_{tot}$, total surface energy $E_{tot}$ and {001} planes occupancy percentage $P_{001}$ (c). In (c) and its insert figure, the volume of the ideal Al cluster is kept 29000 Å$^3$ unchanged.

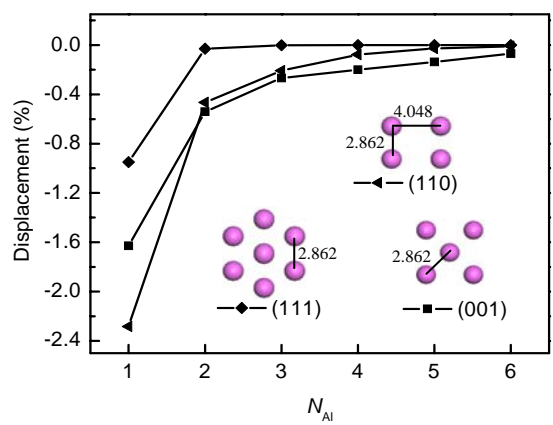

Fig. 1

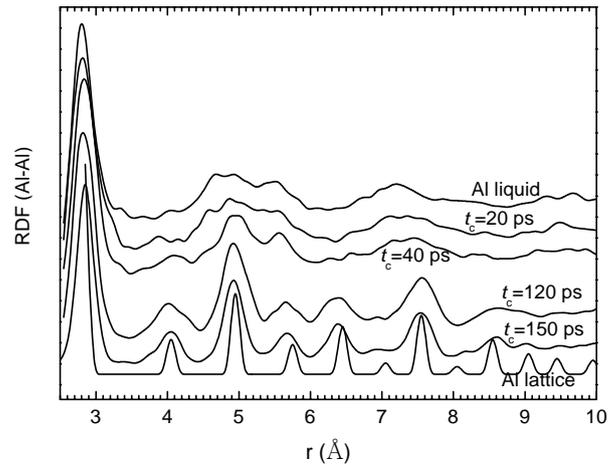

Fig. 2

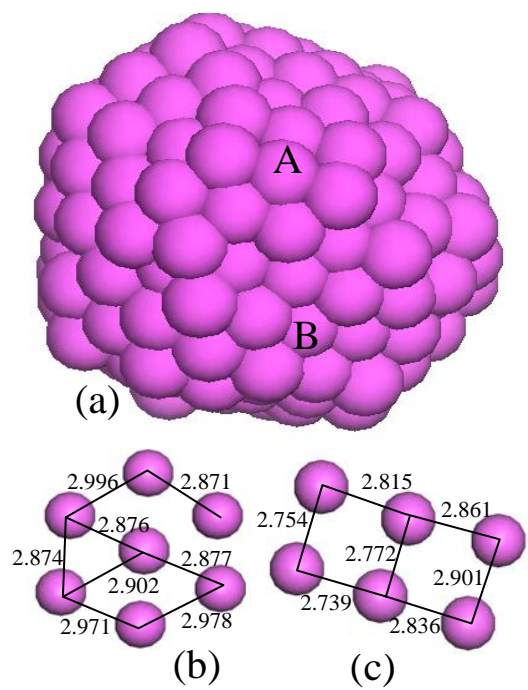

Fig. 3

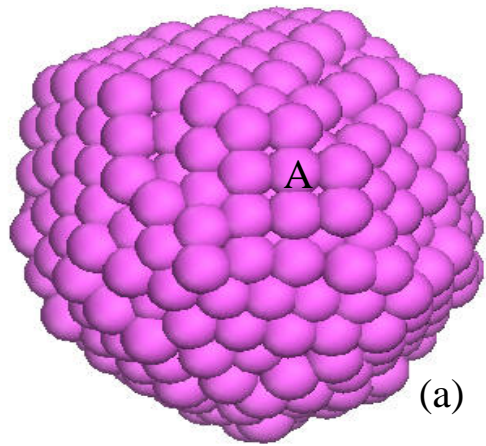
(a)

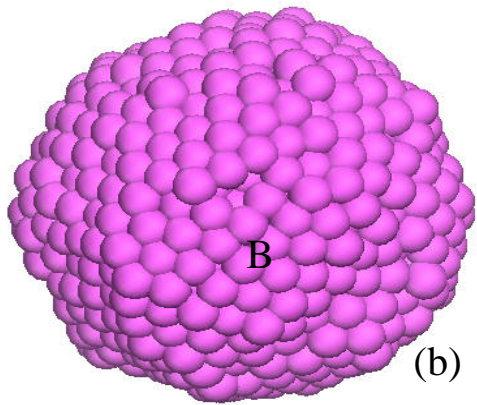
(b)

Fig. 4

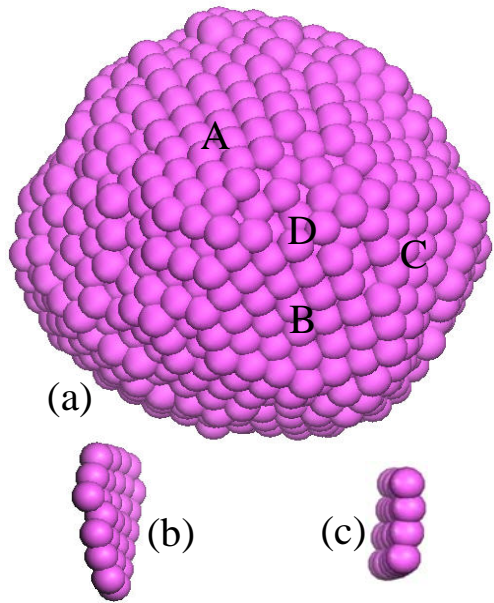

Fig. 5

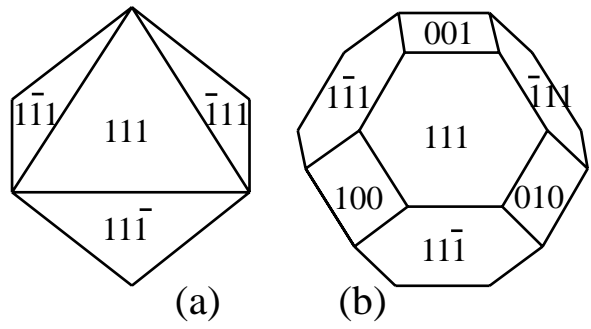
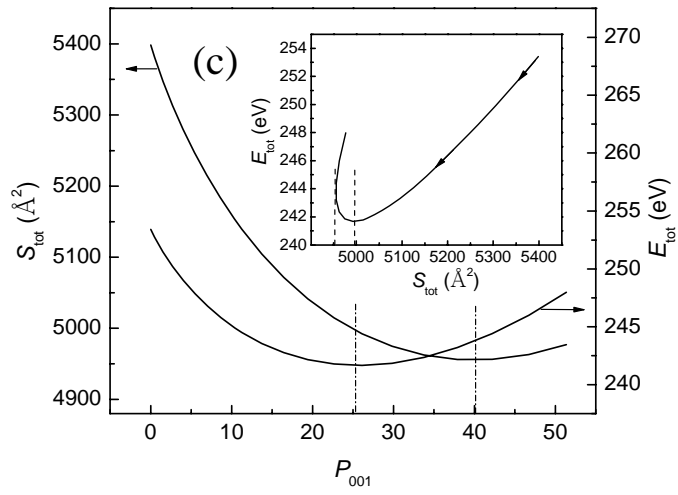

Fig. 6